\documentstyle[aps,floats,12pt]{revtex}
\tighten
\input{epsf}
\begin{document}
\draft
%  
%  Titel and authors
%
\title{Molecular-Dynamics Simulation of a Glassy Polymer Melt:\\
Rouse Model and Cage Effect}
\author{C. Bennemann, J. Baschnagel, W. Paul\footnote{To whom correspondence should be addressed.  
Email: {\sf Wolfgang.Paul@uni-mainz.de}}, and K. Binder\\[2mm]}
\address{Institut f\"ur Physik, Johannes-Gutenberg Universit\"at,\\Staudinger
Weg 7, D-55099 Mainz, Germany}
\maketitle
%
%	Commands for LaTex2e
%
%\newcommand{\mr}[1]{{\mathrm{#1}}}
%
%	Commands for LaTex 2.09
%
\newcommand{\mr}[1]{{\rm #1}}
\renewcommand{\vec}[1]{\mbox{\boldmath$#1$\unboldmath}}
%
%\receipt{ }
%
%  Abstract
%
%       the following lines only necessary if the first
%	documentstyle is used.
%
\bigskip
\begin{center}
{\bf Abstract}
\end{center}
\begin{abstract}
We report results of molecular-dynamics simulations for a glassy 
polymer melt consisting of short, linear bead-spring chains.
The model does not crystallize upon cooling but exhibits a glassy slowing
down. The onset of this slowing down is brought about by the dense packing in
the melt. It was shown in previous work that this onset is compatible with
the predictions of the mode coupling theory of the glass transition. The 
physical process of ``caging'' of a monomer by its spatial neighbors leads to
a distinct two step behavior in scattering functions and particle mean square
displacements. In this work we analyze the effects of this caging process on
the Rouse description of polymer melt dynamics. The Rouse model is known,
both from experimental and simulational work, to be a reasonable description of
the dynamics of short chains in the melt. We show that the Rouse 
description is applicable for length and time scales above the typical
scales for the caging process and that the typical time scale of the Rouse 
model reflects the onset of freezing as described by mode coupling theory.
The Rouse modes are eigenmodes of the chains also in the supercooled state,
and the relaxation times
of the modes exhibit the same temperature dependence as the diffusion
coefficient of the chains. The decay of the mode correlation functions is
stretched and depends on the mode index. Therefore, there is no 
time-mode superposition of the correlation functions. However, they
exhibit a time-temperature superposition at late times. At intermediate
times they decay in two steps for temperatures close to the dynamical 
critical temperature of mode coupling theory.
The monomer displacement is compared with 
simulation results for a binary LJ-mixture to point out the differences
which are introduced by the connectivity of the particles.
\end{abstract}
%
%  Include PACS numbers, and Journal, to which paper will be submitted.
%
\pacs{{\sf PACS}: 61.20.Ja,64.70.Pf,61.25.Hq,83.10.Nn\\
accepted by {\em Comp.\ Theo.\ Poly.\ Sci.}}
%
%  Text of the paper
%
\section{Introduction}
\label{intro}
\noindent
\noindent
In 1953 P. E. Rouse proposed a simple model to describe the 
dynamics of a polymer chain in dilute solution \cite{rouse}.
The model considers the chain as a sequence 
of Brownian particles which are connected by entropic harmonic springs.
Being immersed in a structureless solvent the chain
experiences a random force by the incessant collisions
with the (infinitesimally small) solvent particles. The
random force is assumed to act on each monomer separately and
to create a monomeric friction coefficient. The model therefore
contains chain connectivity, a local friction and a local random
force. All non-local interactions between monomers distant along
the backbone of the chain, such as excluded-volume or hydrodynamic
interactions, are neglected. In dense melts, where both interactions are 
screened, the Rouse model was shown to describe the viscoelastic properties of
short chain melts \cite{ferry,doi,pfgsm}. For the polyethylene melts studied in
\cite{pfgsm} it was also shown by neutron spin-echo experiments \cite{rwzffh,ewen}
and computer simulations \cite{wpmacro,wpprl} that the conformational dynamics
can be reasonably well described by the Rouse model on length scales below the 
radius of gyration and time scales
below the longest relaxation time of the chains (Rouse time). However, by
construction (see section II), it can also only be applicable on length scales
above the statistical segment length, $b$, of the chains and on time scales
larger than $b^2 \zeta/ k_\mr{B} T$, where $\zeta$ is the monomer friction
coefficient. On length scales below the statistical segment length local
stiffness effects due to the intramolecular potentials become important and
there are several suggestions in the literature how to incorporate those into
a modified Rouse model \cite{allegra,winkler}.

For a short chain polymer melt undergoing a glass transition it is also well 
established that the dramatic increase in relaxation times observable for
instance in the viscoelastic response can be accounted for within the Rouse 
model by fitting a temperature dependent monomer friction coefficient in the
form of a Vogel-Fulcher-Tammann (VFT) law \cite{ferry,pfgsm}
\begin{equation}{\label{VFT}
\zeta (T) = \zeta_\infty \exp\left[\frac{E}{T-T_0}\right ] \; .}
\end{equation}
The extrapolated temperature of divergence, $T_0$, in this law (complete structural
arrest) together with the (also extrapolated) vanishing of the excess 
configurational entropy of the glass with respect to the crystal (Kauzmann
paradox \cite{kauz}) lead to theories assuming an underlying phase transition
for the glassy freezing \cite{adam,gdm,freevol}. Especially the 
Gibbs-DiMarzio theory for polymers is capable of reproducing many 
phenomenological properties of the polymer glass transition, the prediction
of a vanishing configurational entropy, however, could be shown to arise from
too crude approximations in calculating the high temperature limit 
\cite{wolfgard}. The physical significance of these extrapolated singularities is 
therefore questionable.

The VFT law has one further characteristic feature. There exists a temperature
region, where the behavior of the viscosity turns over from a gradual high
temperature increase upon lowering the temperature to a very steep increase
in the supercooled region approaching the viscosimetric glass transition 
temperature $T_\mr{g}$. In this temperature region a change in the physical
relaxation mechanism occurs. At high temperatures the mean square displacement
of the particles directly crosses over from short time ballistic to long time
diffusive motion. In this crossover region a plateau regime intervenes where
the particles are temporarily trapped in a cage formed by their neighbors until some
thermally activated process leads to an escape from the cage. Experimentally,
this two step process is best observed in intermediate scattering functions,
and it is well established for all fragile to intermediate glass forming 
systems \cite{ali93,vigo97}, 
i.e., those where the viscosity is describable by the VFT law, but is  
also observable for strong glass formers at very high temperatures
\cite{juergen} (where the viscosity follows an Arrhenius law).

Theoretically it is the mode coupling theory of the glass transition (MCT)
\cite{goetze1,goetze2,yip} that focuses on this temperature region. In its 
idealized version (neglecting the activated processes that lead to an escape
from the cage) it predicts an ergodic to non-ergodic transition at a dynamical
critical temperature $T_\mr{c}$, which phenomenologically seems to be the same
as the inflection point in the VFT law (Fischer et al. in \cite{vigo97}) 
marking the center of the crossover region.

In this work we want to answer the question, how this two step relaxation 
process that is induced by the cage effect influences the Rouse model 
description of the dynamics of short chain polymer melts. 

We have shown in previous work \cite{bpbd,chr_mct,bpbb_pressure} that the
two step relaxation process occurring in our model upon supercooling the 
system is compatible with the MCT predictions and have determined the dynamical
critical temperature $T_\mr{c}=0.45$ and the exponent parameter $\lambda=0.635$
governing the algebraic divergence of correlation times at this temperature
in the idealized version of the MCT. In this work, we will analyze the
conformational relaxation of the chains and their self-diffusion properties
in the same temperature region in terms of the Rouse model.

Section II will introduce our simulation model and a short collection of
results from the Rouse model for later reference. In section III we will then
look at the behavior of the Rouse modes and section IV will focus on the
self-diffusion properties of our model. Section V will present our conclusions.

\section{Simulation Model and Theoretical Background}
In this section we will give a short introduction to our simulation model
and technique and then summarize some pertinent results from the analytic
solution of the Rouse model. A more detailed description of our simulation
model can be found in reference \cite{bpbd}.

\subsection{Simulation Model}
\label{sm}
Our simulation model consists of bead spring chains of length $N=10$. All beads
interact through a Lennard-Jones potential
\begin{equation}{
U_\mr{LJ}(r_{ij}) = 4\epsilon\left[\left(\frac{\sigma}{r_{ij}}\right)^{12} - \left(\frac{\sigma}{r_{ij}}\right)^{6}\right]\, ,}
\end{equation}
which is truncated at $2\times 2^{1/6}\sigma$ and shifted so as to vanish
smoothly at that point.
$\sigma=1$ defines the length scale and $\epsilon=1$ defines the temperature
scale of our model. Bonded neighbors in a chain furthermore interact through
a FENE potential
\begin{equation}{
U_\mr{FENE}(r_{ij}) = -15 R_0^2 \ln\left[ 1 - \Bigg(\frac{r_{ij}}{R_0}\Bigg)^2\right]
}
\end{equation}
with $R_0=1.6$.
The resulting equilibrium bond length is $l_0=0.96$. For a Lennard-Jones potential
that is truncated in the minimum (soft spheres) this is the Kremer-Grest model 
\cite{kg} for which it was shown that the conformational relaxation of short 
melt chains in the high temperature regime can be rather well described by 
the Rouse model \cite{kg,kopf}. Note, however, that the latter model (because
of the strictly repulsive interaction between the monomers) does not yields
physically reasonable equation of state for the polymer melt. Also the friction
coefficient depends on temperature only weakly, unlike equation (\ref{VFT}).

We performed Molecular Dynamics simulations in the canonical (NVT) ensemble 
using the Nos\'{e}--Hoover \cite{nosehoov} thermostat. 
For each temperature, however, the 
equilibrium density is determined by an isobaric-isothermal (NpT) simulation  
before the canonical simulation was started.
In this way we follow a constant pressure path upon cooling 
\cite{bpbb_pressure}.

\subsection{Rouse model}
\label{rousm}
\noindent
The Rouse model is defined through the following equation of motion for the
repeat units of a polymer chain, $\vec{r}_n$ being the position of the $n$-th
effective monomer at time $t$,
\begin{equation}{
\zeta {\rm d}\vec{r}_n(t)= \frac{3 k_{{\rm B}} T}{b^2}\left(\vec{r}_{n+1}(t) - 2\vec{r}_n(t)
+ \vec{r}_{n-1}(t)\right) {\rm d}t + {\rm d}\vec{W}_n(t)\, ,}
\end{equation}
where $b$ is the statistical segment length of the chains and defines the
length scale of the model, $\zeta$ is the monomer friction coefficient and
the ${\rm d}\vec{W}_n(t)$ denote random forces modeled as
Gaussian white noise:
\[\langle {\rm d}W_{n\alpha}(t){\rm d}W_{m\beta}(t')\rangle = \delta_{nm}\delta_{\alpha
\beta}\delta(t-t') {\rm d}t\, .\]
The Rouse modes are defined as the cosine
transforms of position vectors, $\vec{r}_n$, to the monomers. 
For the discrete polymer model under consideration they can be 
written as \cite{verdier}
\begin{equation}
\vec{X}_p(t)=\frac{1}{N}\sum_{n=1}^{N} \vec{r}_n(t)
\cos\left [\frac{(n-1/2)p\pi}{N}\right ] \; , \quad 
p=0,\ldots,N-1 \; .
\label{defx}
\end{equation}
The normalized time-correlation function of the Rouse modes is given by
\begin{equation}
\Phi_{pq}(t)=\frac{\langle \vec{X}_p(t)\vec{X}_q(0)\rangle}{
\langle \vec{X}_p(0)\vec{X}_q(0)\rangle}=
\exp\left [-\frac{t}{\tau_p(T)}\right]\; , \quad
p=1,\ldots,N-1 
\label{defcx}
\end{equation}
with ($p,q\neq 0$)
\begin{equation}
\langle \vec{X}_p(0)\vec{X}_q(0)\rangle=
\frac{b^2}{8N[\sin(p\pi/2N)]^2} \delta_{pq} 
\;\;\stackrel{p/N \ll 1}{\longrightarrow}\;\;
\frac{Nb^2}{2\pi^2p^2}\delta_{pq}
\label{x0}
\end{equation}
and
\begin{equation}
\tau_p(T)=\frac{\zeta(T) b^2}{12 k_\mr{B}T[\sin(p\pi/2N)]^2} 
\;\;\stackrel{p/N \ll 1}{\longrightarrow}\;\;
\frac{\zeta(T) N^2b^2}{3\pi^2k_\mr{B}Tp^2}\; .
\label{tau}
\end{equation}
According to equations (~\ref{defcx}), (\ref{x0}) and 
(\ref{tau}) the Rouse modes should have the following 
properties: (1.) They are orthogonal at all times. (2.)  
Their correlation function decays exponentially. (3.)
The normalized correlation functions for different mode
indices, $p$, and temperatures can be scaled onto a common
master curve when the time axis is divided by $\tau_p(T)$. 
We will examine to what extent these properties are realized by
the studied model. In order to do this comparison we use
the following relations
\begin{equation}
b^2=\frac{R^2}{N-1} \quad \mbox{and} \quad
\frac{\zeta(T) b^2}{k_\mr{B}T}=\frac{R^2}{N(N-1)D} \;,
\label{rouse2lj}
\end{equation}
where $R^2$ ($\simeq 12.3$) is the squared end-to-end distance,
and $D$ is the diffusion coefficient of a chain. The first
relation holds because of the Gaussian chain statistics, which 
is well established in dense melts, and thus fixes the length
scaling between the simulation and the Rouse model. Use of the
second relation means the the diffusive behavior at late times
is employed to fix the translation between the time scale of
the simulation and the theory.

Using equation (\ref{defx}) for the Rouse modes the position vectors
of the monomers can be written as
\begin{equation}
\vec{r}_n(t)=\vec{X}_0(t)+2\sum_{p=1}^{N-1} \vec{X}_p(t)
\cos\left [\frac{(n-1/2)p\pi}{N}\right ] \; , \quad
n=1,\ldots,N \; ,
\label{defr}
\end{equation}
which implies for the mean square displacement of the $n$-th monomer
\begin{equation}
\Big \langle \left [\vec{r}_n(t)-\vec{r}_n(0)\right]^2\Big \rangle =
g_3(t) + 8 \sum_{p=1}^{N-1}\langle |\vec{X}_p(0)|^2\rangle \Big [
1-\Phi_{p}(t) \Big ] \cos^2 \left [\frac{(n-1/2)p\pi}{N}\right ] \; ,
\label{g1.rouse}
\end{equation}
where $g_3(t)=\langle [\vec{R}_\mr{cm}(t)-\vec{R}_\mr{cm}(0)]^2 \rangle$
denotes the mean square displacement of the chains' center of mass.
Note that only the orthogonality of the Rouse modes, i.e., $\Phi_{pq}(t)
\sim \delta_{pq}$, enters the derivation of equation (\ref{g1.rouse}).
\section{Rouse Modes: Time and Temperature Dependence}
\label{results_rm}
\begin{figure}[htb]
\begin{center}
\begin{minipage}[t]{100mm}
\epsfysize=90mm
\epsffile{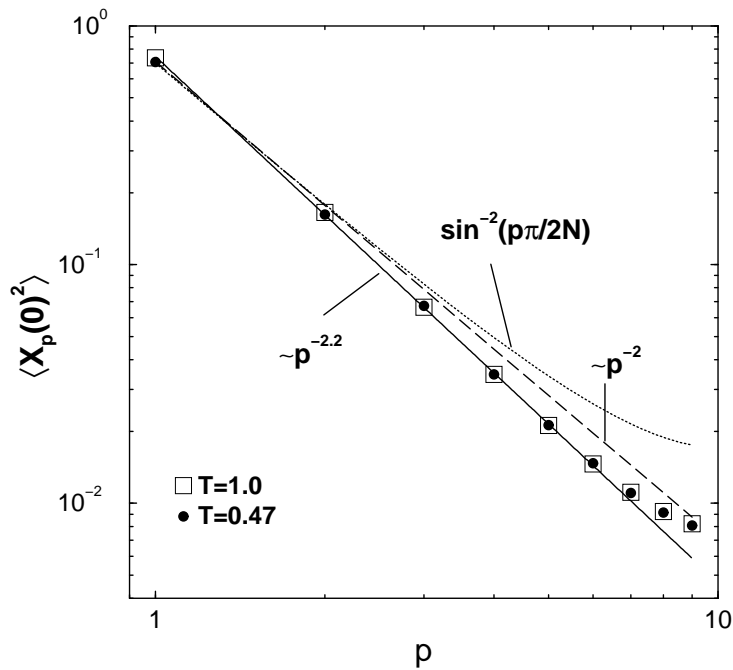}
\end{minipage}
\end{center}
\caption[]{Dependence of the initial correlation of the Rouse modes,
$\langle\vec{X}_p^2(0)\rangle$, on the mode index $p$ at $T=0.47$ and 
$T=1$. The theoretical expectations of the discrete [$\sim \sin^{-2}(p\pi/2N)$]
and continuous ($\sim p^{-2}$) Rouse models are indicated as a dotted and a dashed 
line, respectively. In both cases, the prefactor was not adjusted, but calculated
from the end-to-end distance ($R^2\simeq 12.3$) according to equations (\ref{x0}) and 
(\ref{rouse2lj}). The solid line is a power law fit, $\langle\vec{X}_p^2(0)\rangle 
\sim p^{-x}$, with an effective exponent $x\simeq 2.17$.}
\label{xpstat}
\end{figure}

\noindent
For the Hamiltonian chosen for our simulation we find that the stiffness of
the chains as for instance given by the characteristic ratio only weakly depends
on temperature. We find $C_N=\langle R^2\rangle/(N-1)l_0^2=1.52$ 
for $T=1.0$ and $C_N=1.46$ for $T=0.46$. So
there is a slight shrinking of the chains due to the non-bonded interactions.
The assumption of Gaussian statistics for all intramolecular distances leads
to equation (\ref{defcx}) for the correlation function of the amplitudes of 
the Rouse modes. We find for our model \cite{chr_diss} that the static cross
correlations between different modes $p$ and $q$ are always two to three orders
of magnitude smaller than the static autocorrelation of modes $p$, shown in 
Figure (\ref{xpstat}). 
We can see that the mode amplitudes are independent of
temperature as could be expected from the behavior of the characteristic ratio.
Their qualitative trend as a function of mode number is the same as for the
Rouse prediction of equation (\ref{x0}). Quantitatively, however, already the
second Rouse mode shows a small deviation from the Rouse prediction (full 
curve in Figure (\ref{xpstat}), which was plotted using the calculated value
for the autocorrelation of mode $p=1$ according to equation (\ref{x0}) using
(\ref{rouse2lj}) to relate to quantities measured in the simulation). Up to mode
$p=6$ (this is the scale of a dimer) the behavior can be described by a power
law with an effective exponent $x=2.2$, whereas the small-$p$ expansion of 
equation (\ref{x0}) with exponent $x=2$ fails for $p>3$. These deviations 
from the Rouse prediction stem from the fact that the assumption of Gaussian
distributed intramolecular distances breaks down on the length scale ${R}/{p}$
connected with mode $p$, where $R$ is the end to end distance of the chain.
They depend on the details of the polymer model under investigation \cite{wpmacro,okun}.

\begin{figure}[htb]
\begin{center}
\begin{minipage}[t]{100mm}
\epsfysize=90mm
\epsffile{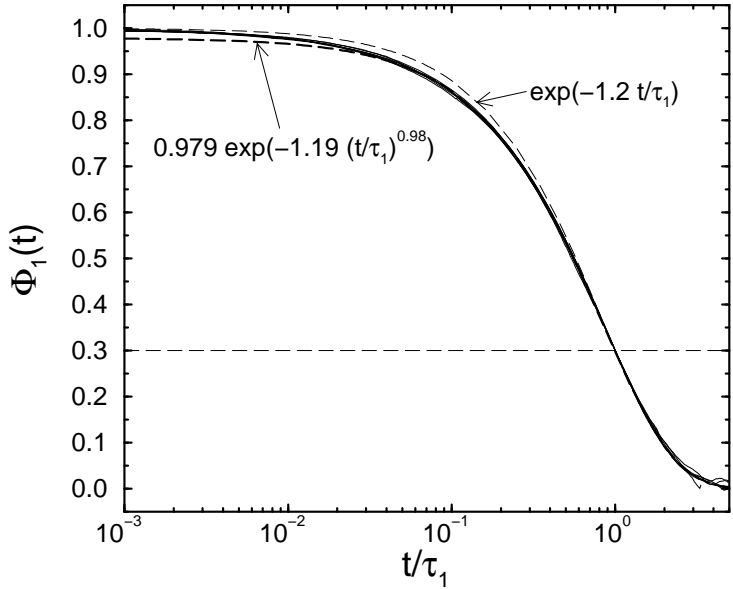}
\end{minipage}
\end{center}
\caption[]{Correlation function of the first Rouse mode, $\Phi_1(t)$, versus rescaled 
time $t/\tau_1$ for seven different temperatures: $T=0.49,0.5,0.52,0.55,0.6,0.65,0.7,1$.
The scaling time $\tau_1$ is defined by $\Phi_1(\tau_1)=0.3$ (dashed horizontal line).
In addition, two exponential functions are shown. The upper (thin dashed line) has an
amplitude of 1. It corresponds to the Rouse prediction [see equation (\ref{defcx})]. The lower
(thick dashed line) is a fit for $0\leq \Phi_1(t) \leq 0.85$.}
\label{cor1}
\end{figure}

Let us now turn to the dynamics of the Rouse modes. In Figure (\ref{cor1}) we
show the dynamic autocorrelation function of the first Rouse mode as a function
of scaled time. The time scale $\tau_1$ is defined as the time value where
the mode autocorrelation function has decayed to a value of 
$\phi_1(\tau_1)=0.3$. According to equation (\ref{defcx}) this should lead to
a master plot showing a single exponential decay, when the autocorrelation 
function for different temperatures are compared. In Figure (\ref{cor1}) we
included all simulated temperatures from $T=0.49$ to $T=1.0$, and we can see
that the time-temperature superposition principle is fulfilled nicely. The
master curve can, however, not be described by a single exponential decay as
is obvious from the figure. If the complete decay is included into a single
exponential fit a law $\exp(-1.2t/\tau_1)$ is obtained. The prefactor of $1.2$
in the argument of the exponential compensates for the $0.3$ definition versus
the $\rm{e}^{-1}$ definition of the relaxation time. If only the last $85\%$
of the decay are included into the Kohlrausch-Williams-Watts fit\
($A_{p}\exp\{-\ln(A_{p}/0.3)(t/\tau_1)^{\beta_{p}}\}$) with the amplitude $A_{p}$ 
and the exponent $\beta_{p}$ as fit parameters, this law gets a prefactor of 
$A_1=0.98$ and an exponent $\beta_1=0.98$, which is equal to one within
our error bars. These two fit curves bracket the observed scaling function for 
scaled times below one. 

\begin{figure}[htb]
\begin{center}
\begin{minipage}[t]{100mm}
\epsfysize=90mm
\epsffile{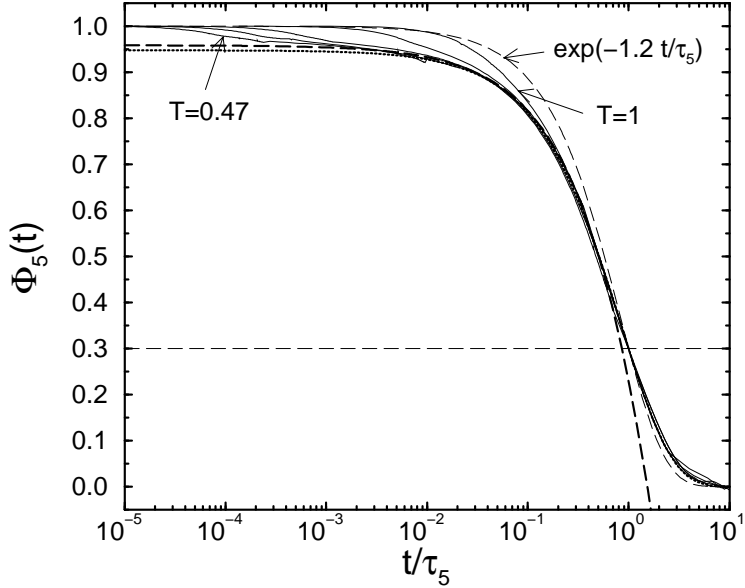}
\end{minipage}
\end{center}
\caption[]{Correlation function of the fifth Rouse mode, $\Phi_5(t)$, versus rescaled 
time $t/\tau_5$ for five different temperatures: $T=0.47$ ,0.49, 0.52, 0.7, 1.
The scaling time $\tau_5$ is defined by $\Phi_5(\tau_5)=0.3$ (dashed horizontal line).
In addition, an exponential function (thin dashed line), a Kohlrausch function (dotted
line), and a von-Schweidler fit (thick dashed line) are shown. The Kohlrausch function
is given by $0.948 \exp[-1.15(t/\tau_5)^{0.87}]$ with $0 \leq \Phi_5(t) \leq 0.95$
as a fit interval. Using the von-Schweidler exponent $b=0.75$ from the mode-coupling 
$\beta$-analysis \cite{chr_mct} and fitting $\Phi_5(t)$ for $10^{-4} \leq t \leq 0.5$, the 
result for the von-Schweidler function is given by $0.959-0.815(t/\tau_5)^{0.75}+0.086
(t/\tau_5)^{1.5}$.}
\label{cor5}
\end{figure}

The reason for the deviation of the scaling function from the
single exponential decay is more obvious when we look at the mode 
autocorrelation function for higher $p$. As an example we show in Figure (\ref{cor5}) the
scaling plot for $p=5$ for $5$ temperatures ranging from $T=0.47$ to $T=1$.
Here the time temperature superposition only works for the late stages of the
decay (scaled times of about $0.5$ and larger), the so-called $\alpha$-process.
The scaling range between the curves at different temperatures, however, 
increases upon lowering the temperature.
The dashed curve in Figure (\ref{cor5}) is the same exponential decay as in
Figure (\ref{cor1}). The master function for $p=5$ is significantly stretched
compared to the single exponential decay. Using again the last $85\%$ of the
decay for a fit to the Kohlrausch-Williams-Watts law, we get a 
stretching exponent of $\beta_5=0.81$ 
(dotted line). This stretching exponent for the $\alpha$-relaxation is, 
however, independent of temperature in the shown temperature interval, which
covers the high temperature liquid region ($T=1.0$) as well as the supercooled
fluid region ($T=0.47$). This is the same result as was found in \cite{okun} 
for a Monte Carlo simulation of a polymer lattice model undergoing a glass
transition, in contrast to the discussion in \cite{loring}. The decrease in
the Kohlrausch exponent $\beta_p$ upon supercooling that is predicted in that
model calculation is not observed in our simulation. The stretching, however,
strongly depends on mode number and the values for the exponent $\beta_p$ for
the different Rouse modes are collected in Table $1$. This mode number
dependence was also found in the lattice simulation \cite{okun} as well as an
atomistic simulation of a polyethylene melt \cite{wpmacro,wpprl}. 

For scaled times below about $0.5$, the $\alpha$-scaling breaks down. For $p=5$ 
we can resolve the development of a two step decay for temperatures $T\le0.7$
with an intervening plateau that increases in length upon cooling. This is the
manifestation of the cage effect in the Rouse modes. The consequences of this
caging for the structural relaxation behavior of our model was analyzed in
detail in \cite{chr_mct}. It was not observable for $p=1$ because the amplitude
of the plateau is too close to one in that case (even for $p=9$ it is still
larger than $0.9$), but it leads to the discussed deviations from the single
exponential decay. This plateau regime is called $\beta$-relaxation within
mode coupling theory and the decay off the plateau should be describable by
a von Schweidler-like law 
\begin{equation}{\label{vons}
\Phi_5\left(\frac{t}{\tau_5}\right)=f_5^{\rm c} - B_1 \left(\frac{t}{\tau_5}\right)^b
+ B_2\left(\frac{t}{\tau_5}\right)^{2b} \; ,}
\end{equation}
where we included the first correction to the leading order in analogy
to the predicted behavior of the incoherent scattering function \cite{fuchs}. 
We take the
von Schweidler exponent $b=0.75$ from our previous work \cite{chr_mct} and
fit the von Schweidler law to the data in the time interval $10^{-4}\le
{t}/{\tau_5}\le 0.5$ and obtain a very good description of the asymptotic
scaling in this $\beta$-regime with the parameters $f_5^{\rm c}, B_1$ and $B_2$ quoted
in the caption of Figure (\ref{cor5}). 

Because of the mode number dependent stretching a time-mode superposition cannot be 
\begin{figure}[htb]
\begin{center}
\begin{minipage}[t]{100mm}
\epsfysize=80mm
\epsffile{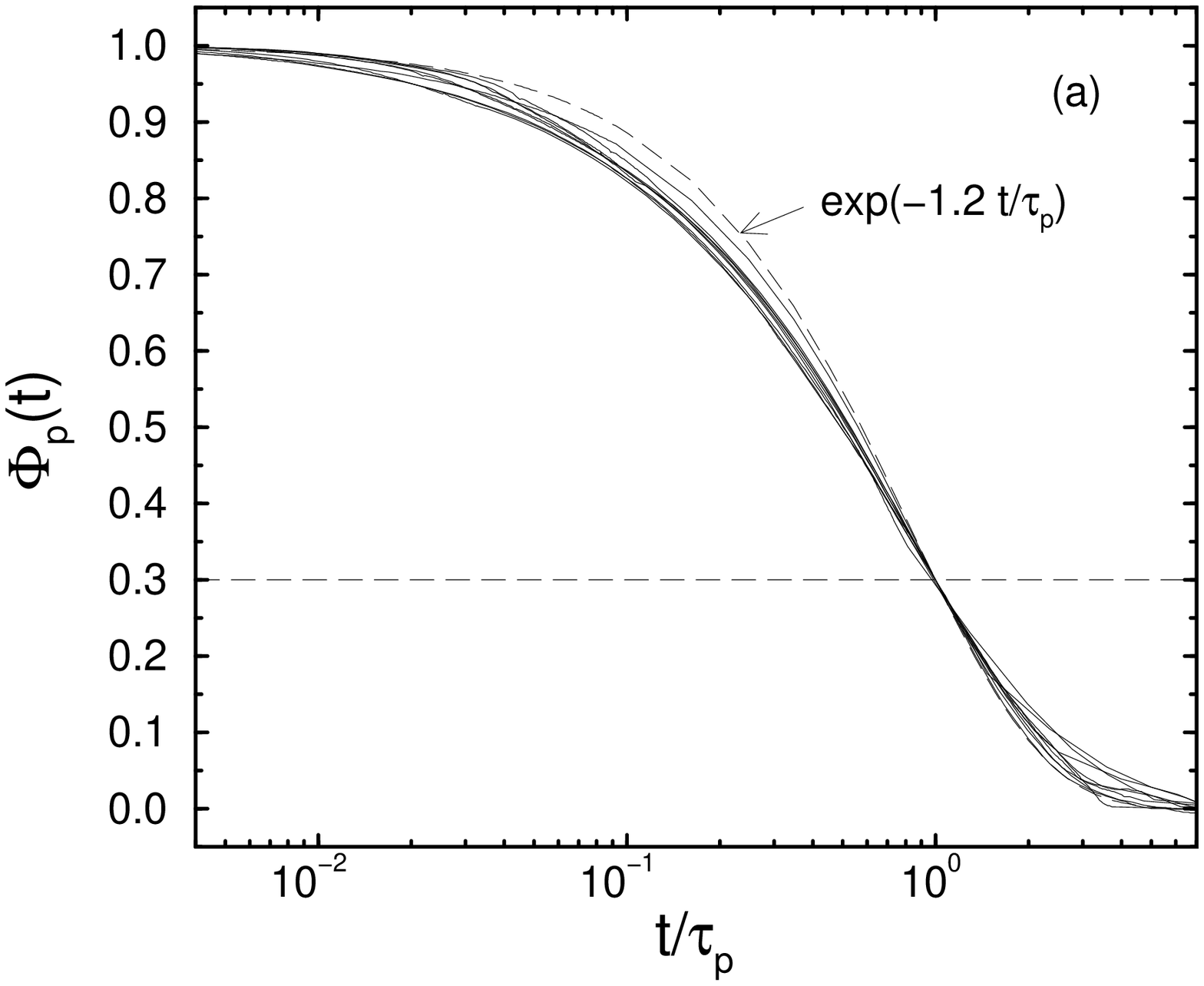}
\vspace{-8mm}
\epsfysize=80mm
\epsffile{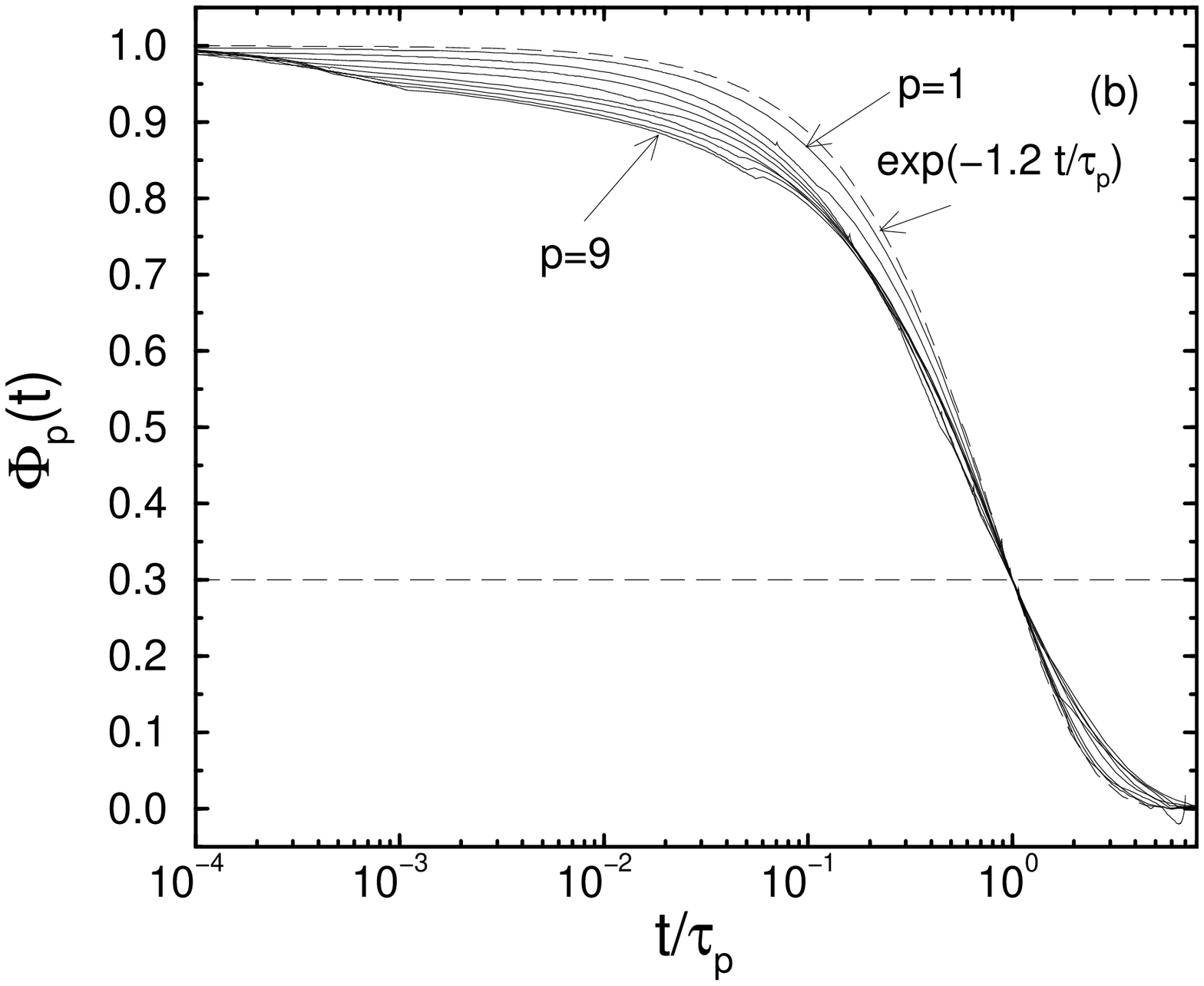}
\end{minipage}
\end{center}
\vspace{-8mm}
\caption[]{Correlation functions of all nine Rouse modes, $\Phi_p(t)$, versus rescaled 
time $t/\tau_p$ for two different temperatures: $T=1$ [normal liquid state of the melt;
panel (a)], $T=0.49$ [supercooled state close to $T_\mr{c}=0.45$ \cite{chr_mct}; panel
(b)]. The scaling time $\tau_p$ is defined by $\Phi_p(\tau_p)=0.3$ (dashed horizontal 
line). In addition, an exponential function (thin dashed line) is shown.}
\label{corp}
\end{figure}
expected to hold. In Figure (\ref{corp}a) we show the failure of the
time-mode superposition for the high temperature liquid case ($T=1$) and
Figure (\ref{corp}b) shows the same for the supercooled state $(T=0.49)$. 
Here we can also observe the development of the $\beta$-plateau with increasing
mode number with a plateau value of about $0.9$ for $p=9$. 

\begin{figure}
\begin{center}
\begin{minipage}[t]{100mm}
\epsfysize=90mm
\epsffile{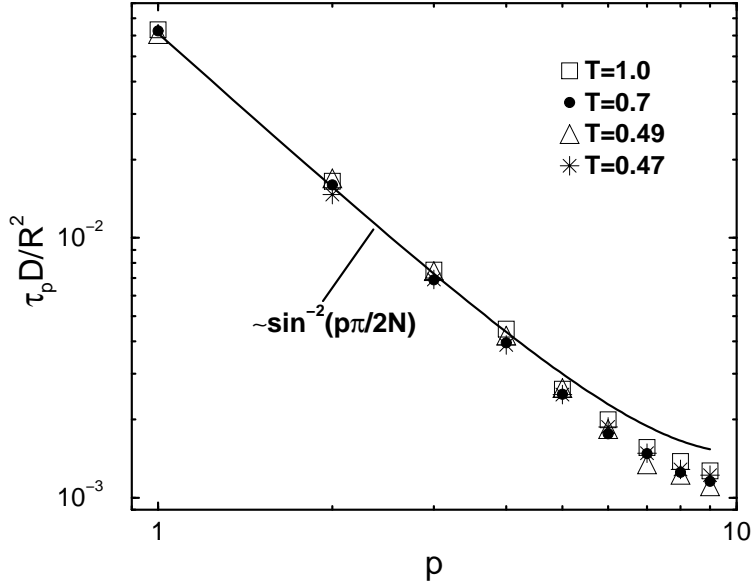}
\end{minipage}
\end{center}
\caption[]{Variation of the Rouse mode relaxation time with the mode index $p$ for 
four different temperatures: $T=0.47,0.49,0.7,1$. The relaxation time $\tau_p$ is 
defined as the time value, at which the correlation function of the Rouse modes
has decayed to 0.3, i.e., $\Phi_{pp}(\tau_p)=0.3$. If $T<0.49$, the decay of the
function for the first mode is too slow to reach 0.3 within the 
simulation time. Therefore, only the results for $p \geq 2$ are shown for $T=0.47$.
As anticipated by equations (\ref{tau}) and (\ref{rouse2lj}), the temperature dependence
of $\tau_p$ scales as $1/D$ ($R$ is essentially temperature independent). The
solid line is the prediction of the discrete Rouse model for the $p$-dependence of
$\tau_p$. The prefactor is $1.62 \times 1/12N(N-1)$. Contrary to 
Figure \ref{xpstat}, an additional factor (i.e., 1.62) was necessary to
shift the Rouse prediction onto the simulation data because the correlation functions
do not exhibit a simple exponential decay so that the used definition of $\tau_p$
($\Phi_{pp} (\tau_p)=0.3$) can deviate from the Rouse definition ($\Phi_{pp} (\tau_p)=
\mr{e}^{-1}$) by a numerical constant.}
\label{tp}
\end{figure}

When we look at the temperature dependence of the mode relaxation times
$\tau_p$ within the Rouse model, equations (\ref{x0}) with (\ref{rouse2lj})
tell us that the quantity $\tau_p D / R^2$, where $D$ is the center of mass
diffusion coefficient and $R^2$ the squared end-to-end distance of the chains,
should be independent of temperature. Figure (\ref{tp}) shows that this is
indeed the case and that the dependence on mode index can be very well 
described by the Rouse prediction up to $p=4$. The improved agreement with the
Rouse model in comparison to the static 
behavior in Figure (\ref{xpstat})
suggests that the deviations from the conformational assumptions in the Rouse
model are partially compensated by deviations from the dynamic assumptions 
going into the model, as they are manifest in the stretching of the modes.

\begin{figure}
\begin{center}
\begin{minipage}[t]{100mm}
\epsfysize=90mm
\epsffile{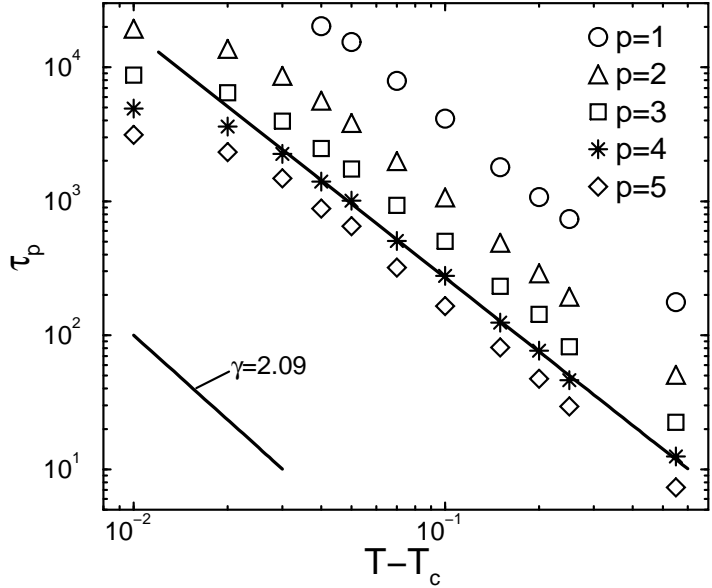}
\end{minipage}
\end{center}
\caption[]{Log-log plot of the Rouse relaxation time $\tau_p$ versus $T-T_\mr{c}$ for
the first five Rouse modes.  $\tau_p$ is defined by $\Phi_{pp}(\tau_p)=0.3$, and 
$T_\mr{c}=0.45$ is the critical temperature of mode-coupling theory (MCT) \cite{chr_mct}. 
Additionally, two solid lines are shown: The lower one indicates the exponent expected 
from the MCT $\beta$-analysis, i.e., $\gamma=2.09$ \cite{chr_mct}. The other
line through the data point for $p=4$ represents a fit result for $0.49 \leq T \leq
1$, yielding $\gamma_p=1.83 \pm 0.02$. Within the error bars, this exponent provides a
reasonable fit not only for $p=4$, but for all modes shown.}
\label{tp.mct}
\end{figure}

In Figure (\ref{tp.mct}) we finally analyze the temperature dependence of the
mode relaxation times within the framework of the mode coupling theory of the
glass transition. To this end, we plot the measured relaxation times 
double-logarithmically as a function of the distance to the mode coupling critical
temperature $T_\mr{c}=0.45$ as determined for our model in \cite{chr_mct}. For about
one decade in the reduced temperature we observe an algebraic behavior as 
predicted for the $\alpha$-relaxation time within MCT. For $T\le0.49$ 
deviations from the algebraic divergence at $T_\mr{c}$ occur, where ergodicity 
restoring processes that were neglected in the idealized MCT calculation start
to become important. The exponent we observe is $\gamma_p=1.83\pm0.02$, which
is within the error bars equal to the exponent observed for the temperature
dependence of the self-diffusion coefficient of the chains $\gamma_D\simeq 1.82$. 
The error bar for $\gamma_p$ indicates the scattering between the fits to the
different modes when fixing 
$T_\mr{c}=0.45$. The exponent is distinctly different from the 
$\gamma=2.09$ obtained from a $\beta$-analysis of the incoherent scattering 
function in \cite{chr_mct}. The agreement of the temperature dependence of an
orientational correlation time as given by $\tau_1$ and a translational
correlation time definable by $R^2/D$ was also found experimentally from a
comparison of dielectric and pulsed-field gradient NMR data \cite{schoenhals}.
It also gave rise to the scaling observed in Figure (\ref{tp}). We can 
conclude from Figure (\ref{tp.mct}) that all modes show a freezing transition
at the same temperature $T_\mr{c}=0.45$ and with the same exponent $\gamma_p$
in contrast to an analysis in \cite{vakhtan} which, however, had to employ a
frozen matrix assumption. If one treats matrix and probe chain dynamics 
self-consistently a unique freezing temperature is again obtained \cite{rehkopf}.

In order to interpret the difference in exponent between the incoherent 
scattering function and the conformational relaxation times of a chain as
given by the Rouse modes, we have to keep in mind that the relaxation times
at the mode coupling $T_{\rm c}$ do not actually diverge but stay finite. When we
compare the temperature dependence of $\tau_q(T) / \tau_q(T_\mr{c})$ for a given
momentum transfer $q$ with the temperature dependence of
$\tau_p(T) / \tau_p(T_\mr{c})$ for a given Rouse mode $p$, the latter quantity
decays much slower with increasing distance $T-T_\mr{c}$ due to the presence of
connectivity correlations between monomers of the same chain.

\begin{figure}[htb]
\begin{center}
\begin{minipage}[t]{100mm}
\epsfysize=90mm
\epsffile{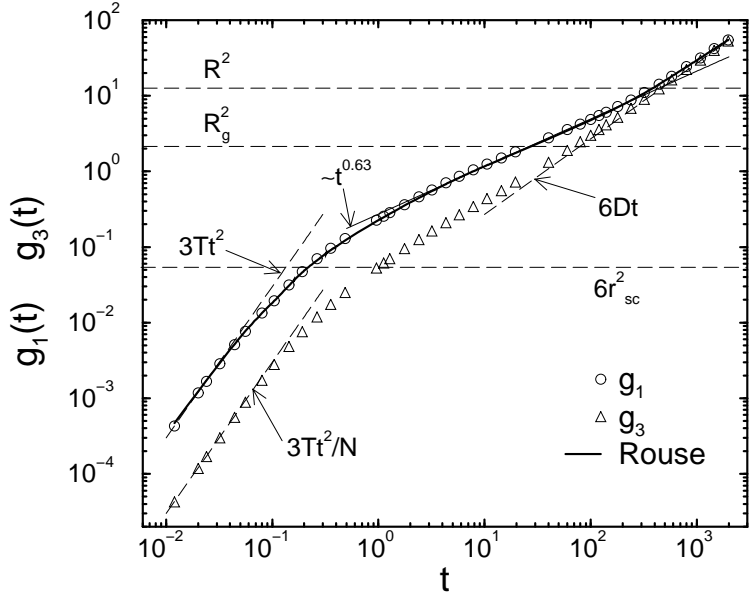}
\end{minipage}
\end{center}
\caption[]{Log-log plot of the mean-square displacements of an inner monomer, $g_1(t)$, 
and of the chains' center of mass, $g_3(t)$, versus time for $T=1$. The thick solid
line is the Rouse formula for $g_1(t)$ [see Eq.~(\ref{g1.rouse})]. The initial ballistic
behaviors for $g_1(t)$ and $g_3(t)$, i.e., $g_1(t)=\langle v^2\rangle t^2=3Tt^2$ and 
$g_3(t)=3Tt^2/N$ ($\langle v^2\rangle$: mean-square monomer velocity, $N=10$: chain length),
and the late time diffusive behavior are indicated as dashed lines. In addition, a power
law fit $g_1(t) \sim t^x$ with an effective exponent $x \simeq 0.63$ is shown as a thin solid
line. The dashed horizontal lines represent the end-to-end distance $R^2$ ($\simeq 12.3$; upper 
line) and the radius of gyration $R_\mr{g}^2$ ($\simeq 2.09$; lower line), respectively.} 
\label{g1.T=1}
\end{figure}

\section{Rouse Model and Mean Square Displacements}
\label{msds}
\noindent

In this section we want to analyze the translational behavior of central 
monomers of a chain, $g_1(t)=\langle [\vec{r}_{N/2}(t)-\vec{r}_{N/2}(0)]^2\rangle$,
 and of the center of mass of a chain, $g_3(t)$,
as a function of temperature. Figure (\ref{g1.T=1}) displays these functions in
the high temperature case ($T=1$). For short times both displacements are 
ballistic and equal to $\vec{v}^2 t^2=3Tt^2$ and $\vec{v}^2 t^2/N$, respectively,
in reduced units where $k_\mr{B}=1$ and the monomer mass is set to one. Then we see
a crossover to a subdiffusive behavior in $g_1$ which is induced by the 
connectivity of the chains (Rouse mode dominated regime). The exponent in this
regime is $0.63$ instead of the Rouse prediction of $0.5$, which is a deviation
generally found in simulations and mostly attributed to the shortness of the
chains leading to an early crossover from the Rouse mode regime to the long
time free diffusion limit. This last regime is seen to occur, when the mean 
square displacements of the monomers are equal to the squared end-to-end
distance of the chains. The center of mass displacement $g_3$ reaches the
free diffusion limit at earlier times, namely when it is equal to the mean
squared radius of gyration of the chains. The full line in Figure 
(\ref{g1.T=1}) proves in another way that the Rouse modes are eigenmodes of the
chains. 
This curve is calculated from the mode autocorrelation functions using
equation (\ref{g1.rouse}). As a 
\begin{figure}
\begin{center}
\begin{minipage}[t]{100mm}
\epsfysize=90mm
\epsffile{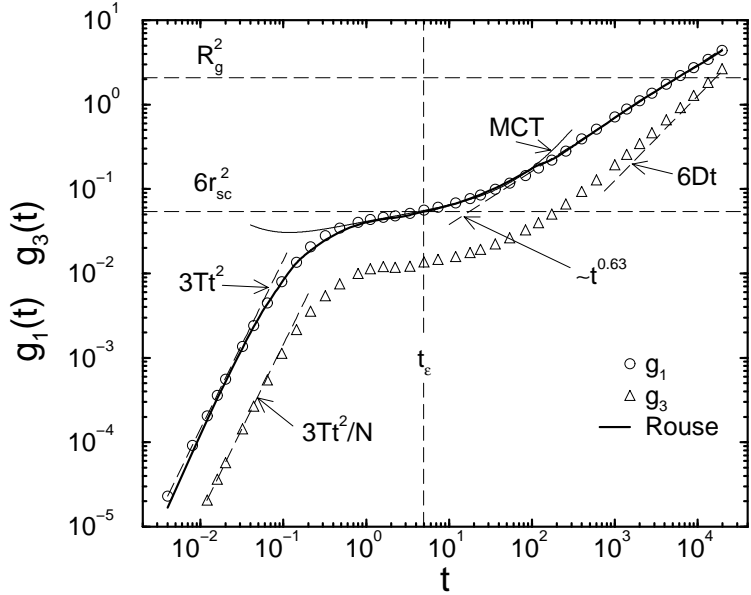}
\end{minipage}
\end{center}
\caption[]{Same as Fig.~\ref{g1.T=1}, but for $T=0.48$. Additionally, the mode-coupling
approximation in the $\beta$-relaxation regime, i.e., equation (\ref{msdmct}), is reproduced from
Ref.~\cite{chr_mct}. The dashed horizontal lines indicate the plateau value, $6r_\mr{sc}^2$,
of equation (\ref{msdmct}) (lower line), the radius of gyration $R_\mr{g}^2$ ($\simeq 2.09$; middle 
line) and the end-to-end distance $R^2$ ($\simeq 12.3$; upper line),
respectively. The dashed vertical line indicates the $\beta$ time scale $t_\varepsilon$
from equation (\ref{msdmct}).} 
\label{g1.T=0.48}
\end{figure}
consistency check we exactly reproduce the monomer mean square displacement curve. We have 
to use all Rouse modes to
obtain this exact agreement. If one only wants to describe the time dependence
for $g_1(t)\ge R_g^2$, the first five Rouse modes suffice for our model.
Also included in the Figure is a horizontal line
at $6 r_\mr{sc}^2$ indicating the size of the next neighbor cage as obtained in
\cite{chr_mct}. For this displacement value we observe at high temperatures 
the crossover from the short-time ballistic motion to the short time diffusion
and further the onset of the connectivity dominated regime. Both occur at the
same distance scale, as the length scale of the non-bonded interaction, $\sigma$, 
and the bond length, $l_0$, are approximately equal.

For the supercooled melt this picture changes qualitatively, as can be seen in
Figure (\ref{g1.T=0.48}). For this temperature of $T=0.48$ we were only able 
to propagate the chains for about two squared radii of gyration (after equilibration for
about one order of magnitude longer in time). In the
supercooled melt a plateau-like regime which extends for this temperature for
about two decades in time intervenes between the short time ballistic regime
and the connectivity dominated regime where the subdiffusive behavior again
shows the exponent $0.63$. This plateau is at the value of the cage size and
it is centered on the $\beta$-relaxation time scale $t_\varepsilon$ of MCT
\cite{chr_mct}. The MCT prediction for the mean squared monomer displacement 
$\langle (\Delta r)^2\rangle$, averaged over all monomers,
for the $\beta$-regime is
\begin{figure}[htb]
\begin{center}
\begin{minipage}[t]{90mm}
\epsfysize=90mm
\epsffile{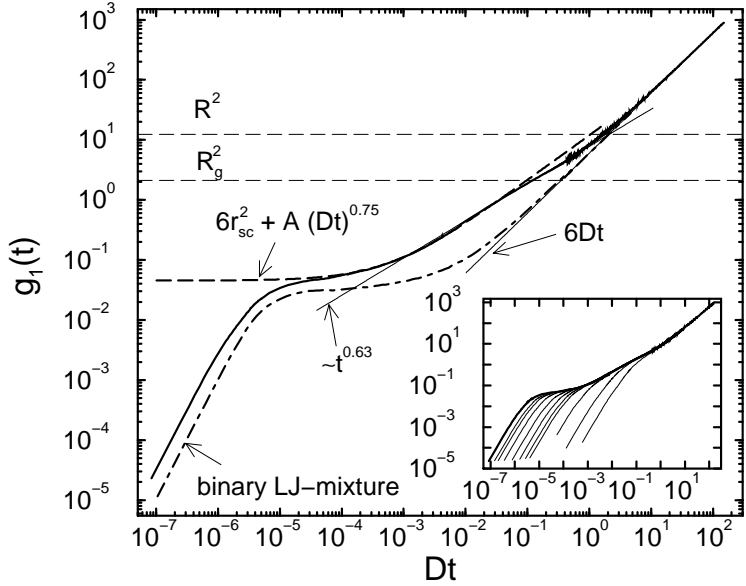}
\end{minipage}
\end{center}
\vspace{-8mm}
\caption[]{Master curve of the mean-square displacement of an inner monomer $g_1(t)$ 
versus rescaled time $Dt$ ($D$: diffusion coefficient of a chain). The master curve (= thick
solid line in the inset and the main figure) is constructed from the temperatures $T= 0.48$, 0.49,
0.5, 0.52, 0.55, 0.6, 0.65, 0.7, 1, 2, 4 (from left to right in inset; the labels of the inset axes are 
identical to those of the main figure). The thick dash-dotted line is a master curve constructed
from the simulation data of a binary Lennard-Jones mixture \cite{ka1}, including temperatures
from $0.466 \leq T \leq 5$. Since $T_\mr{c} \simeq 0.435$ for the binary mixture, but $T_\mr{c} 
\simeq 0.45$ for the present polymer model, the lowest temperatures correspond to the same 
distance to the critical point, i.e., $T-T_\mr{c}=0.03$, in both cases. Additionally, two thin
straight lines are shown, indicating a power law fit to the monomer displacement, $g_1(t) \sim t^x$
with $x \simeq 0.63$, and the late time diffusive behavior $6Dt$ which is common to the simple
liquid and the polymer data. The thick dashed line is a fit to an effective 
von Schweidler law (see text).
The dashed horizontal lines represent the end-to-end distance $R^2$ ($\simeq 12.3$; upper line) and 
the radius of gyration $R_\mr{g}^2$ ($\simeq 2.09$; lower line), respectively.} 
\label{g1.Dt}
\end{figure}
\begin{equation}{\label{msdmct}
g_1(t)\approx\langle (\Delta r)^2\rangle = 
6r_\mr{sc}^2-6h_\mr{msd}\left [\frac{t_0}{t_\varepsilon}\right]^{a} g(t/t_\varepsilon)
-6h_\mr{msd}C_a\left [ \frac{t_0}{t}\right]^{2a}
-6h_\mr{msd}B^2C_b\left [ \frac{t_0}{t_\varepsilon}\right]^{2a}
\left [ \frac{t}{t_\varepsilon}\right]^{2b}\; .}
\end{equation}
The parameters $r_\mr{sc}^2=0.009$, $h_\mr{msd}t_0^a=0.0045$, 
$t_\varepsilon=4.933$, $a=0.352$, $b=0.75$, $B=0.476$
$C_{\it{a}}t_0^a=-0.3$ and $C_{\it{b}}t_0^a=-0.25$  
are taken from reference \cite{chr_mct}.
As was already discussed in that work, the MCT prediction for the $\beta$-regime
for our model allows for a consistent description of the cage effect on the 
mean square monomer displacement. The MCT curve would predict a crossover from
the breakup of the cage to the free diffusion of the particles, as no 
connectivity effects are included in the theory. For a polymer fluid, however,
this behavior is altered and we obtain a crossover to the mode dominated regime,
$t^{0.63}$, which is subdiffusive with a smaller exponent than the cage breakup
described by the von Schweidler $t^{0.75}$ law (in leading order). 

Figure 
(\ref{g1.T=0.48}) tells us also why an analysis of the caging process in a polymer
melt within the framework of a theory developed for simple liquids can work at
all. The typical distance traveled by a monomer at the plateau is of the order of 
$10^{-1}$ in units of the Lennard-Jones length scale which in turn is approximately equal 
to the bond length in our model. For this length scale connectivity effects are 
not yet felt by the monomers even in the high temperature regime shown in 
Figure (\ref{g1.T=1}). Only the late stage of the cage breakup (late $\beta$-process)
and the structural relaxation ($\alpha$-process) are influenced by the 
connectivity of the chains. It is also noteworthy that the analysis in terms of
the Rouse modes works throughout the cage region as can be concluded from the
full line in Figure (\ref{g1.T=0.48}), obtained in the same way as for Figure 
(\ref{g1.T=1}). Furthermore the caging process is not only observed in the 
monomer mean square displacement, but in the displacement of the center of 
mass of the chains as well.

\begin{figure}
\begin{center}
\begin{minipage}[t]{100mm}
\epsfysize=90mm
\epsffile{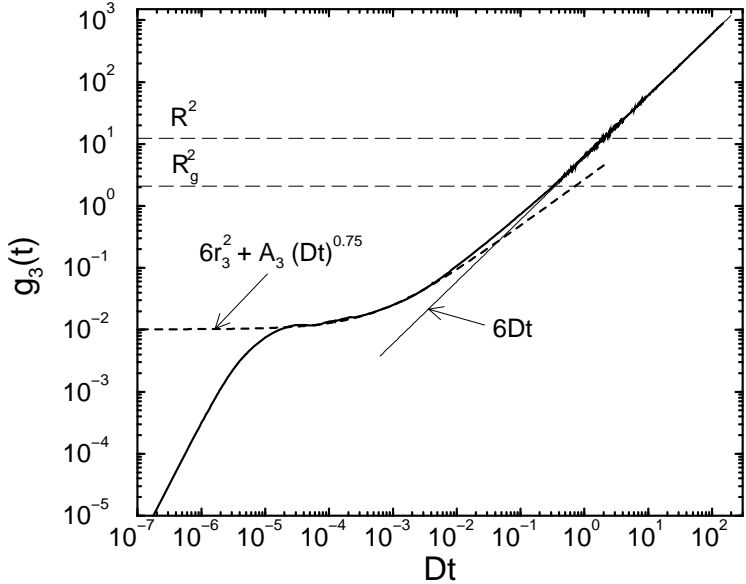}
\end{minipage}
\end{center}
\caption[]{Same as in Fig.~\ref{g1.Dt}, but for the mean-square displacement of the chains'
center of mass $g_3(t)$. The master curve is constructed from the same temperatures as used
for $g_1(t)$. The thin solid line indicates the late time diffusive behavior $6Dt$, and the 
thick dashed line is again a fit to an effective von Schweidler law (see text).
The dashed horizontal lines represent the end-to-end distance $R^2$ ($\simeq 12.3$; upper line) and 
the radius of gyration $R_\mr{g}^2$ ($\simeq 2.09$; lower line), respectively.} 
\label{g3.Dt}
\end{figure}

The difference between a polymer melt and a simple liquid in the $\beta - \alpha$ 
crossover region is elucidated from a different angle in Figure (\ref{g1.Dt}).
If we plot the mean square displacement for all temperatures as a function of
time scaled by the center of mass diffusion coefficient at that temperature we 
obtain the set of curves displayed in the inset of Figure (\ref{g1.Dt}).
The envelope master curve of this set of curves is shown in the main part of
Figure (\ref{g1.Dt}) in comparison to the same master curve constructed for
a binary Lennard-Jones fluid \cite{ka1,kob_rev95}. For large times the data for the two
models have to agree by construction. But whereas for the Lennard-Jones fluid
a direct crossover from the cage effect to the free diffusion occurs, the polymer
exhibits the intervening connectivity dominated regime for length scales between
the bond length, $l\approx 1$, and the end-to-end distance $R$. In this regime the
observed mean square displacement curve drops below the MCT description, which
is here displayed as an effective von Schweidler law $6r_1^2 + A_1(Dt)^{0.75}$,
with $r_1=0.087$ and $A_1=11.86$. The same type of fit is possible for the
displacement of the center of mass of the chains as is shown in Figure (\ref{g3.Dt}).
The center of mass of a chain exhibits qualitatively the same crossover form the
cage regime to the free diffusion regime as seen for the simple liquids. The fit
values for the effective von Schweidler law in this case are $r_3=0.041$ and 
$A_3=2.68$.

\section{Conclusions}
\label{conc}
\noindent 
We have shown in this work that the Rouse modes stay eigenmodes of the 
dynamics of short melt chains from the high temperature to the supercooled
fluid state. They are orthogonal and for the smallest mode numbers they obey
the predictions of the Rouse model for their static amplitudes. When the length
scale probed by a mode is too short to allow for Gaussian distributed 
intramolecular distances on that scale, deviations from the static and
dynamic scaling predictions occur. 
Nevertheless, when one uses the actual Rouse mode correlation functions to 
predict the time-dependent mean square displacements, one
obtains perfect agreement with the simulation results.

The most important deviation from the Rouse model prediction is a mode number
dependent stretching of the time autocorrelation function of the mode 
amplitude. This stretching is within our accuracy not temperature dependent 
for temperatures above the mode coupling critical temperature of our model,
which were accessible for a dynamic analysis starting from equilibrated melts.
We therefore have to conclude that it is not so much connected with the 
approach to the glass transition, but with deviations from the simple Rouse
model due to 
intramolecular stiffness and intermolecular interaction effects.

For temperatures in the supercooled fluid region the mode autocorrelation
functions develop a clear two step decay similar to the behavior of the 
intermediate scattering function as discussed in \cite{chr_mct} for the same model. 
The decay in the plateau region is compatible with the $\beta$-analysis 
following MCT presented earlier. Furthermore, the mode relaxation times
follow the $\alpha$-scaling behavior of MCT with the same dynamic critical
temperature obtained from the intermediate scattering function. The exponent
of the seeming algebraic divergence is the same as for the center of mass 
self-diffusion coefficient of the chains and different from the exponent seen
in the intermediate scattering function. The difference can be understood in
terms of additional correlations between monomers connected along the same
chain, that are not included in the MCT, which was developed for simple
liquids. 

This connectivity leads to the Rouse prediction of a subdiffusive monomer
displacement at intermediate times between the short time ballistic regime
and the long time free diffusion limit. In the supercooled melt the cage
effect, that leads to the discussed two step decays, intervenes between the 
short time ballistic motion of a monomer and the subdiffusive Rouse behavior. 
Therefore there is no direct crossover from the cage breakup to a free diffusion
behavior as for simple liquids, but a crossover to the connectivity dominated
regime with a smaller exponent than the von Schweidler exponent, which 
describes the cage breakup. Since the higher Rouse modes do not contribute to
the  center of mass displacement (which is the zeroth Rouse mode), the center
of mass of the chains exhibits qualitatively the same crossover from cage
region to free diffusion as observed  for simple liquids.

The whole applicability of the MCT to the supercooled polymer melt rests on
the fact that the caging process occurs on a length scale of about one tenth
of the bond length. On this scale the monomers just start to feel the connectivity
to their neighbors and therefore only the late stages of the cage breakup and
the structural relaxation are affected by the connectivity of the chains. The
applicability of the Rouse model starts for length scales larger than the
bond length and time scales of the order of the structural relaxation time
of the melt.
%
%  Acknowledgment
%
\section*{Acknowledgment}
We are indebted to Drs.\ W. Kob, M. Fuchs
and I. Alig for helpful discussions.
In the course of this work, we have profited
from generous grants of simulation time by the computer
center at the university of Mainz and the HLRZ J\"ulich,
which are gratefully acknowledged, as well as financial support 
by the Deutsche Forschungsgemeinschaft under SFB262/D2.
%
%  References
%

%
%	Tables
%
\begin{table}[t]
\begin{tabular}{l|l|l}
$p$ & $A_p $& $\beta_p$\\\hline
$1$ & $0.99$ & $0.98$ \\[0.2cm]
$2$ & $0.96$ & $0.92$ \\[0.2cm]
$3$ & $0.96$ & $0.84$ \\[0.2cm]
$4$ & $0.96$ & $0.82$ \\[0.2cm]
$5$ & $0.97$ & $0.81$ \\[0.2cm]
$6$ & $0.98$ & $0.78$ \\[0.2cm]
$7$ & $1.00$ & $0.77$ \\[0.2cm]
$8$ & $0.97$ & $0.77$ \\[0.2cm]
$9$ & $0.94$ & $0.83$ \\[0.2cm]
\end{tabular}
\vspace{3mm}
\caption{KWW amplitude $A_p$ and stretching exponent $\beta_p$ as a function of
mode number. The KWW function was fitted to the last $85\%$ of the decay.
There is a systematic decrease of $\beta_p$ with increasing mode number $p$, 
i.e. a systematic stretching of the modes. The exponent value for for $p=9$ 
seems to deviate from the trend, but also the KWW fit became worse for this
high mode number.}
\end{table}
\end{document}